# Issues and Challenges in Applications of Artificial Intelligence to Nuclear Medicine - The Bethesda Report (AI Summit 2022)

Arman Rahmim
Departments of Radiology and Physics, University of British Columbia

Tyler J. Bradshaw
Department of Radiology, University of Wisconsin - Madison

Irène Buvat
Institut Curie, Université PSL, Inserm, Université Paris-Saclay, Orsay, France

Joyita Dutta
Department of Biomedical Engineering, University of Massachusetts Amherst

Abhinav K. Jha
Department of Biomedical Engineering and Mallinckrodt Institute of Radiology, Washington University in St. Louis

Paul E. Kinahan
Department of Radiology, University of Washington

Quanzheng Li
Department of Radiology, Massachusetts General Hospital and Harvard Medical School

Chi Liu
Department of Radiology and Biomedical Imaging, Yale University

Melissa D. McCradden
Department of Bioethics, The Hospital for Sick Children, Toronto

Babak Saboury
Department of Radiology and Imaging Sciences, Clinical Center, National Institutes of Health

Eliot Siegel
Department of Radiology and Nuclear Medicine, University of Maryland Medical Center, USA

John J. Sunderland
Departments of Radiology and Physics, University of Iowa

Richard L. Wahl
Mallinckrodt Institute of Radiology, Washington University in St. Louis

**Abstract**

The SNMMI Artificial Intelligence (SNMMI-AI) Summit, organized by the SNMMI AI Task Force, took place in Bethesda, MD on March 21-22, 2022. It brought together various community members and stakeholders from academia, healthcare, industry, patient representatives, and government (NIH, FDA), and considered various key themes to envision and facilitate a bright future for routine, trustworthy use of AI in nuclear medicine. In what follows, essential issues, challenges, controversies and findings emphasized in the meeting are summarized.

**Introduction**

The SNMMI Artificial Intelligence (SNMMI-AI) Summit, organized by the SNMMI AI Task Force, took place in Bethesda, MD on March 21-22, 2022. As summarized in Fig. 1, various community members and stakeholders from academia, healthcare, industry, patient representatives, and government (NIH, FDA) participated in the AI Summit; and the meeting included rich presentations, roundtable discussion and interactions on key themes to envision and facilitate a bright future for routine, trustworthy use of AI in nuclear medicine. The summit included:

1) Plenary lectures on the role of ethics in AI, state-of-the-art in AI, AIM-AHEAD (AI/ML Consortium to Advance Health Equity and Researcher Diversity), as well as implementation science.

2) Lively panel discussions on perspectives from data scientists, industry representatives (pharmaceuticals, AI software developers, device companies), end-users of AI (physicians, technologists, hospital administrators), and government agencies (NIH and FDA).

3) A call to action.

The meeting, having been the first in-person meeting for many since the pandemic, was considered a great success. In what follows, essential issues, challenges, controversies, and conclusions highlighted during the meeting, discussions and conclusions are outlined.

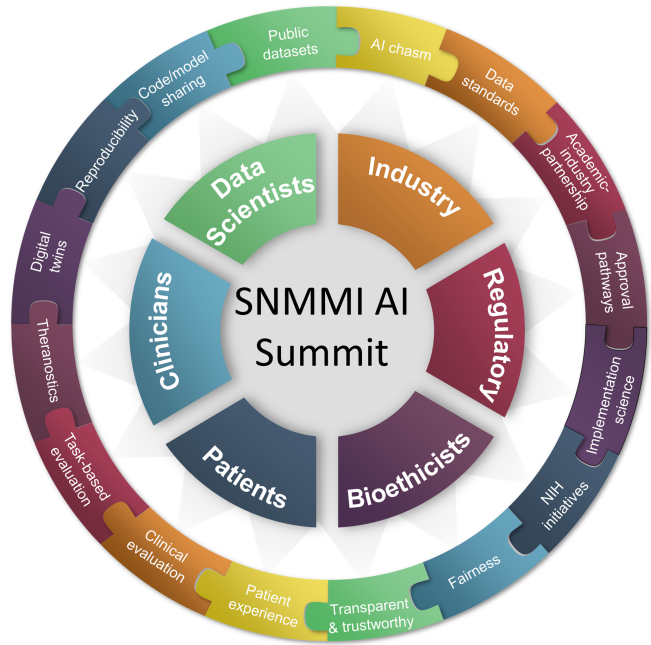

Fig. 1: The SNMMI AI Summit, and key participants/stakeholders and themes.

**Key Frontiers in Translational AI**

Role of Ethics in AI:
1) A framework for model evaluation must include more than just technical considerations. There are multiple dimensions that contribute to clinical efficacy, including fairness, ethics, human-computer interactions, among others. These methods of evidence-gathering are best aligned with the informational needs of bedside clinicians using AI.
2) We should be cautious about AI explainability methods, as explanations can persuade physicians into following incorrect suggestions. Inaccurate advice from AI systems might degrade a physician's diagnostic performance [1]. AI-informed decisions should be explained to patients, and nuclear medicine staff should be educated to understand the potential but also the limitations of AI-based tools, and use them accordingly.
3) Fairness in AI goes beyond the technical performance of a given algorithm. Measuring performance across patient demographic and clinical subgroups is an important first step toward promoting fairness, but must be complemented with an implementation strategy that prioritizes the intended impact to equity-seeking groups.
4) The publication and use of AI-solutions should not transgress the well-known pillars of medical ethics, so new AI-based methods with too little evidence of effectiveness should not be published.
5) Given the many opportunities and associated challenges that AI introduces in the Nuclear Medicine field, priorities should be established to enable and promote a safe and beneficial use of AI solutions.
6) More details appear in the Appendix.

Bridging the AI Chasm in Nuclear Medicine:

1) AI is going to enter all aspects of nuclear medicine [2] and will: 1) enhance image quality and accuracy while reducing scanning time and dose; 2) assist interpretation by automating labor-intensive or operator-dependent tasks; 3) reveal new molecular and metabolic mechanisms and introduce new (diagnostic) biomarkers or predictive/prognostic models; 4) accelerate workflows and data integration while optimizing patient management (see Figure 2).
2) Yet, there is currently a large gap between the expectations associated with AI and the actual services provided by AI in nuclear medicine departments. This gap has been called the AI chasm. The reasons for this gap are many, as described below, but solutions exist to harvest the potential of AI in our field.
3) High-quality and contextualized "AI ready" databases of nuclear medicine images including annotation are needed. Relatively few large-scale nuclear medicine datasets exist, and few data science challenges focused on nuclear medicine have been organized [3,4]. Building them will require a collective effort given that much fewer nuclear medicine images are acquired than for other modalities. This could potentially be accomplished through federated learning platforms. Data efficient approaches to training such as transfer learning, weak-, and self-supervised algorithms could also be more taken advantage of.
4) Large-scale multi-center evaluation of AI-based solutions are scarce and urgently needed. For example, a network of centers could be organized to validate well-documented AI-based software independent of the developers, and evaluation results, including failures, would be systematically reported in a public register.
5) For some applications, such as predicting patient outcomes, explainable AI models should be considered so that the user could check and trust the model [5]. Additionally, models should generate uncertainties with their predictions, and there should be clear guidance on how to resolve conflicts between the algorithm and physician decisions.
6) Easy-to-deploy and well-validated solutions should be shared as much as possible to enhance the quality of care, but also to reduce the gap between institutions that can afford high-end AI technologies and others.
7) Continual learning strategies should be implemented to progressively enhance deployed AI algorithms. Additionally, ongoing maintenance is needed to detect drift in performance possibly associated with subtle changes, for instance in patient population or imaging protocols.
8) Harmonization across societies might be needed to make all initiatives converge faster.

Implementation science in translational AI:
1) Implementation science is the study of methods to facilitate the integration of research findings and evidence into healthcare policy and practice [6].
2) In general, diffusion of innovation in medicine is slow. On average, implementation of the medical research result into routine clinical practice takes 17 years [7].
3) Furthermore, the impact of a technology characteristically decreases from the initial efficacy trial to the actual implementation ("voltage drop") reference [8].
4) We have to consider implementation and dissemination (D&I) earlier, by focusing on questions like: who's going to deliver it? Is it fit to intended patient population? What are the unintended consequences of implementation? What are the impediments and how to facilitate the process?
5) We should not assume that systems are static, patients are homogeneous, and choosing to not implement is irrational. Dynamism should be embraced.
6) The integration of values into technology can enable the delivery of efficacious, socially-licensed AI that improves health outcomes and maintains trust in healthcare systems.
7) A conceptual model of evidence based practice implementation in public service includes: exploration, adaptation decision/preparation, active implementation, and sustainment [9].

Theranostic Digital Twins:
1) Digital twins for precision oncology present a major frontier for precision cancer care [10].
2) The theranostics paradigm is very aptly fit for developing digital twins for our patients; i.e. creating digital models built for individual patients based on theranostics imaging (and other images/data) towards virtual therapies for optimal selection of therapies to be delivered to individual patients (e.g. optimal injected radioactivites, injection profiles, etc.)
3) Emerging applications on radiopharmaceutical therapies (RPTs) are presently commonly practiced in the paradigm of fixed injected radioactivities for patients.
4) There are two major technological impediments to personalization of RPT: disconnect between micro-scale dosimetry and macro-scale dosimetry; and lack of personalized physiologically based pharmacokinetic (PK) modeling frameworks. Neural networks informed by biological models [11] have the potential to use limited data to personalize models for individual patients.
5) Theranostic digital twins may be constructed for individual patients towards optimization of RPTs. This construct combines the methods of multi-scale modeling with nuances of PBPK frameworks. Theranostic Digital Twinning is a platform to integrate multi-modality treatment measures (e.g. chemotherapy, radiotherapy, and RPT) in order to capture heterogeneity, plan interventions, and monitor the response. This is the vision of the future of RPT that includes personalization based on pre- and/or intra-therapy molecular (and other) imaging.
6) Different injection strategies and intervals can be explored. In addition, corrective strategies such as adaptive dose planning, or adjuvant therapy with locoregional therapy (e.g. ablative therapy or external radiation therapy), or systemic chemotherapeutic strategies (e.g.immunotherapy or CAR-T therapy) can be explored.

**Perspectives from the various stakeholders in AI**

Data scientists:
1) The panelists identified the scarcity of high-quality labeled datasets as a primary barrier to the development of AI algorithms. Barriers to translating AI include technical pitfalls, such as poor reproducibility and poor generalizability, but can also include logistical challenges in integration with hospital infrastructures.

2) Rigorous evaluation studies are crucial for clinical translation of AI algorithms. The SNMMI AI Task Force has proposed in a recent report [REF - Jha] that AI algorithms should be evaluated based on clinical tasks to avoid misleading interpretations [12]. The team proposed that an evaluation study should lead to a quantifiable claim, and they put forth a four-class framework for evaluation studies. Physicians have a crucial role in performing these evaluation studies [13].
3) It was recognized that we must consider the interoperability of AI algorithms across different software systems (e.g., PACS) especially when operating on data with different formats (e.g., DICOM, raw, listmode).
4) Guidelines for developing [2], evaluating and reporting AI algorithm performance are now available [14].

Industry (pharma, AI software developers, device companies, Academic-Industry Partnerships):
1) The industry panel was attended by both startup companies as well as large established companies who shared their perspectives on AI for nuclear medicine.
2) A perspective on Academic-Industry Partnerships was also offered from the viewpoint of experienced investigators from academia.
3) Reasons *why* participation in Academic-Industry Partnerships is beneficial were noted, as well as the unique and productive advantage that has been enjoyed by researchers in nuclear medicine to date.
4) Both industry and academic institutions see the value of AI in nuclear medicine and are actively pursuing projects and clinical translation of these projects.
5) General issues with Academic-Industry Partnerships related to AI were noted:
    a. From the viewpoint of industry, each academic center has its own processes for sharing data or algorithms.
    b. From the viewpoint of academia, each company has its own processes for sharing data or algorithms.
    c. The above leads to delays and excessive work. There was discussion of how a unified approach would alleviate these challenges.
6) Unique to nuclear medicine, as opposed to medical imaging in general:
    a. There is a long-standing tradition of close collaboration in Academic-Industry Partnerships
    b. AI algorithms need large amounts of data for training, and independent cohorts for validation. This is a challenge for medical imaging in general, but is even more challenging for nuclear medicine, where imaging volumes are much lower nationally and globally.
    c. For image-reconstruction, image-enhancement and image-segmentation applications, the low resolution and high noise in nuclear-medicine images compared to other anatomical imaging modalities poses challenges for defining ground truth.
    d. The NIH-funded Medical Imaging and Data Resource Center (MIDRC, www.midrc.org) repository was raised as an example of pooling and curating de-identified DICOM images from many centers to enable AI research on larger collections of image data.

e. The experience of MIDRC has shown the need for thoughtful de-identification and common standards for descriptions of imaging studies, as well as the need for curation. I.e. We do not always need 'high-quality' data, as many centers will provide 'standard-of-care' data, rather we need data with *measured* quality.
    f. Establishing universal standards/formats for raw data storage is considered critical for scaling up deep learning efforts for image reconstruction in the nuclear medicine domain. Industrial and academia must work hand in hand toward this goal. This is potentially more feasible in nuclear medicine than medical imaging in general.
7) Improvements in image quality do not always translate to improvement in image statistics and even further down improvements in clinical outcomes. This has led to a persistent gap that remains to be addressed by image generation and processing software.

End users of AI (Physicians, technologists, hospital administrators):
A point that received significant attention was the importance of patient and caregiver perspectives. Recent studies are showing that patients are interested in the prospects of AI as a decision-making adjunct for physicians. Further, patients want their values to be incorporated in the decision making process [15]. This becomes especially relevant in the context that AI algorithms often output uncertainty in their decisions. Methods to incorporate this uncertainty are an important research frontier.

NIH and FDA:
1) The strategic plan of NIH for data science envisions a "modernized, integrated, and FAIR biomedical data ecosystem".
2) Open access data sharing is a priority.
3) To enhance the biomedical data-science research workforce through improved programs and novel partnerships, STRIDES (The **S**cience and **T**echnology **R**esearch **I**nfrastructure for **D**iscovery, **E**xperimentation, and **S**ustainability) has been initiated.
4) NIH Cloud Platform Interoperatability (NCPI) aims to establish and implement guidelines and technical standards to create a federated data ecosystem. Using this, diverse users can co-analyze data to drive science.
5) Health equity and researcher diversity are important topics
6) To enhance the participation and representation of underrepresented community in development of AI model AIM-AHEAD consortium is formed (Artificial Intelligence/Machine Learning Consortium to Advance Health Equity and Researcher Diversity)
7) Although AIM-AHEAD has focused on electronic health record (EHR) in the begining, the scope includes other diverse data such as medical imaging
8) To prepare AI-ready data sets, Bridge2AI initiative aims to use biomedical grand challenges. The process of data set preparation should instill a culture of ethical inquiry.
9) " Achieving the effective convergence of biomedical data and machine learning requires datasets to be thoughtfully designed from the outset to be valuable for machine learning-based analysis." –NIH ACD Working Group Report

10) AI-readiness should be guided by a concern for human and clinical impact.
11) NIBIB funding opportunities include bioengineering partnership with industry (BPI, U01), national centers for biomedical imaging and bioengineering (NCBIB, P41), as well as other individual funding mechanisms (R01, R21, etc). "NIH Katz Award" is a new R01 FOA, released on 11/09/2020, specifically for early-stage investigators focusing on innovations that represents a change in research direction (high risk/high reward projects).
12) FDA objective is to protect and promote public health. Safety and effectiveness of medical devices are the main concerns of FDA's Center for Devices and Radiological Health (CDRH).
13) Risk-based approach to device classification determines the degree of regulation necessary (Class I: low risk; Class II: moderate risk, Class III: high risk device which requires premarket approval)
14) most devices with AI in the nuclear medicine diagnostic space are anticipated to be Class II devices require 510k notification
15) 510(k) premarket notifications are applicable when there is a predicate device with same intended use and same technology, OR same intended use and different technology without different safety/effectiveness concerns.
16) "Intended Use" is the major characteristics of "tool type" claims.
17) Image quality evaluation is "task-specific"
18) FDA emphasizes transparency in the context of AI based on feedback from the community.

**Conclusions and Call to Action**

The summit identifies significant need for:
1) Creation of an AI Center of Excellence to help facilitate an ecosystem for trustworthy AI, in collaboration with numerous collaborators and stakeholders (health systems, nuclear medicine providers, patients, payors, regulatory agencies, industry, etc.)
2) Extensive collection and curation of datasets for benchmarking AI models for image reconstruction and processing
3) Furthering research on federated approaches to facilitate model sharing to circumvent clinical data sharing hurdles
4) Integration of uncertainty modeling and reporting into image generation models in a way that does not burden radiologists or confuse patients

**Acknowledgements**
We wish to give special thanks to Bonnie Clarke with significant efforts to organize AI task force meetings as well as the 2022 AI Summit.

**Appendix**

Background for the SNMMI Task Force

The AI task force of SNMMI was formed in the summer of 2020. Comprised of nearly 20 members, including nuclear medicine physicists and physicians, engineers, computational imaging scientists, statisticians, and representatives from industry & regulatory agencies, the task force has aimed to make recommendations, educate, and enable new opportunities and capabilities towards effective integration of AI in nuclear medicine and molecular imaging, including translation to routine clinical practice. Fig. 2 depicts a cycle (patient-to-image and image-to-which) as a conceptual paradigm along which AI is expected to make significant contributions, and based on which, numerous discussions in the AI task force have taken place. So far, the task force has created 4 main reports (published [2] [14] or undergoing journal review), on:

1) Opportunities, challenges, and responsibilities in nuclear medicine towards creating trustworthy AI ecosystems: This report aims to help establish and maintain leadership in AI by envisioning and motivating concerted efforts to promote the rational and safe deployment of AI in nuclear medicine, including effective engagement of all stakeholders.

2) Best practice for algorithm development: This document aims to educate the community and to provide guidelines on best practices in order to avoid key pitfalls of AI. We have made general recommendations, followed by descriptions on how one might practice these principles for specific topics within nuclear medicine.

3) Best practices for algorithm evaluation: Specifically, we have introduced the RELAINCE (Recommendations for Evaluation of AI for Nuclear Medicine) guidelines. The goal of this work is to provide best practices to evaluate AI algorithms for different objectives including for proof of concept, technical task-specific efficacy, clinical efficacy, and post-deployment effectiveness, in general, and in specific contexts of nuclear medicine imaging.

4) Ethical considerations in point-of-care integration of AI: important points of ethical reflection for AI include balancing innovation with the moral obligation to collect rigorous evidence to support clinical use. Specifically, viewing these considerations through the lens of health disparities and inequity points us to a positive obligation to test AI for performance disparities that produce inequitable risk distribution within patient populations. Prospective clinical testing further contributes to the evidence base that aligns with current evidence-based practice and provides the kind of information that is most valuable to clinicians at the point-of-care. Finally, we explored the practices and provocations of explainability practices (the application of secondary computational methods to elucidate the algorithmic logic behind a given prediction). These include a recognition that clinicians desire understanding behind why an AI system produced a given output, but the need to balance this curiosity against potential exacerbation of automation basis and the unreliability of current explainability methodologies. Developing values-based guidance for the development, testing, and integration of AI systems into nuclear medicine can mitigate potential harms and deliver responsible, efficacious technologies to the point-of-care.

The above four efforts are being promoted in our community via conference educational abstracts and organized continuing education (CE) sessions, as well as journal publications.

Furthermore, the valuable discussions in the task force had a direct impact on the structure and approach by editors of two special AI issues (2021, 2022) by the journal PET Clinics, including contributions by many members of the SNMMI AI task force, as directly acknowledged by the journal. In addition, we had representation in a panel towards a "guideline on radiomics in nuclear medicine", jointly sponsored by the European Association of Nuclear Medicine (EANM) and SNMMI.

Finally, two new initiatives are currently underway by the task force:

1) Data infrastructures: AI methods tend to require significant data for training, challenged by data sharing limitations. Our task force seeks to enable improved and effective methods of AI algorithm learning, including possibilities of creating large, centralized databases as well as organization of multiple data challenges, and frameworks for effective model sharing. We are presently working closely with an NSF-sponsored initiative: "Leveraging Data Communities to Advance Open Science" to help make informed decisions in these important directions.

2) Linking AI and dosimetry towards precision radiopharmaceutical therapy (pRPT): there is increasing enthusiasm in the community towards the significant potential of personalized dosimetry for improved treatment of patients. Nonetheless, dosimetry is often perceived as difficult, cumbersome, and/or in need of improved reproducibility. We believe AI can make significant contributions towards automated and reliable dosimetry applications. This can in turn accelerate implementation of dosimetry-based treatment optimization in routine clinical practice.

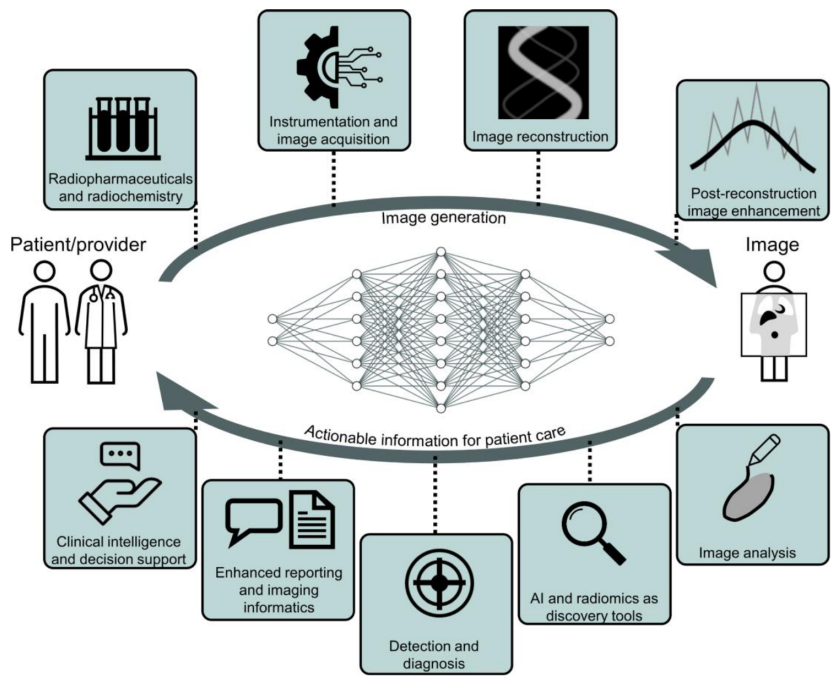

Fig. 2: Applications of AI span the gamut of nuclear medicine subspecialties (based on our published report [2]).